\documentclass[prl,twocolumn,showpacs,preprintnumbers,amsmath,amssymb]{revtex4}

\usepackage{epsfig}
\usepackage{graphicx}
\usepackage{dcolumn}
\usepackage{bm}

\newcommand{\EQ}{\begin{equation}}
\newcommand{\EN}{\end{equation}}
\newcommand{\ea}{\end{eqnarray}}
\newcommand{\ba}{\begin{eqnarray}}

\newcommand{\bear}{\begin{eqnarray}}
\newcommand{\ear}{\end{eqnarray}}

\begin{document}

\title{Quantum criticality and correlations in the cuprate superconductors}
\author{J. M. P. Carmelo} 
\affiliation{GCEP-Centre of Physics, University of Minho, Campus Gualtar, P-4710-057 Braga, Portugal}
\date{12 June 2010}


\begin{abstract}
A description of the electronic correlations contained in the Hubbard model 
on the square-lattice perturbed by very weak three-dimensional uniaxial anisotropy 
in terms of the residual interactions of charge $c$ fermions and spin-neutral composite two-spinon 
$s1$ fermions is used to access further information on the origin of 
quantum critical behavior in the hole-doped cuprate superconductors. Excellent quantitative agreement with  
their anisotropic linear-$\omega$ one-electron scattering rate and normal-state linear-$T$ resistivity 
is achieved. Our results provide strong evidence that the normal-state linear-$T$ resistivity is indeed a 
manifestation of low-temperature scale-invariant physics.
\end{abstract}

\maketitle

The interplay between quantum critical behavior \cite{LSCO-1/lam-x,critical,critical-2,Resistivity,QPT}
and the mechanism underlying the pairing state of the high-temperature superconductors 
\cite{ARPES-review,2D-MIT,duality} remains an enigmatic issue.  
The Hubbard model on the square lattice \cite{ARPES-review,2D-MIT,symmetry,companion2,cuprates0} 
perturbed by very weak three-dimensional uniaxial anisotropy provides the simplest realistic  
description of the role of correlations effects in the properties of the hole-doped cuprate superconductors. 
Recent experiments on these systems
\cite{critical-2,Resistivity,two-gaps,Yu-09,k-r-spaces,PSI-ANI-07,two-gap-Bi,Hm,LSCO-ARPES-peaks,scattering-rate} 
impose new severe constraints on the mechanisms responsible for their unusual properties. 

The virtual-electron pair quantum liquid (VEPQL) \cite{cuprates0} describes 
the above toy model electronic correlations in terms of residual
$c$ - $s1$ fermion interactions. Alike the Fermi-liquid 
quasi-particle momenta \cite{Pines}, those of the $c$ and $s1$ fermions 
are close to good quantum numbers \cite{companion2,cuprates0}. 
The results of Ref. \cite{cuprates0} provide evidence that 
for a hole concentration domain the VEPQL short-range spin order coexists with a long-range $d$-wave superconducting order
consistent with unconventional superconductivity being mediated by magnetic fluctuations \cite{Yu-09}.
The $U(1)$ phase symmetry broken below $T_c$ refers to the hidden $U(1)$ symmetry recently 
found in Ref. \cite{symmetry}. Each  {\it virtual-electron pair} configuration involves one $c$ fermion pair
of charge $-2e$ and one spin-singlet two-spinon $s1$ fermion 
whose spin-$1/2$ spinons are confined within it. 
\begin{figure}
\includegraphics[width=5.0cm,height=3.575cm]{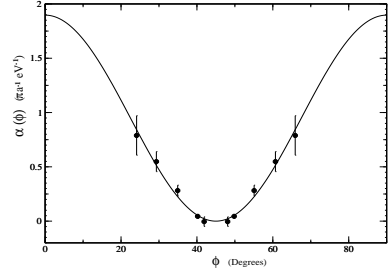}
\caption{The theoretical coefficient $\alpha (\phi) = (\cos 2\phi)^2/(x\,64x_*\sqrt{\pi x_{op}}\,t)$
(solid line) for $x=0.145$ and the LSCO parameters $x_*=0.27$ and $x_{op}=0.16$ together 
with the experimental points for the corresponding coefficient $[\alpha_I (\phi)-\alpha_I (\pi/4)]$ 
of Fig. 4 (c) of Ref. \cite{PSI-ANI-07}.}
\label{fig1} 
\end{figure}

The magnitudes of the basic parameters appropriate to YBa$_2$Cu$_3$O$_{6+\delta}$ (YBCO 123) and 
La$_{2-x}$Sr$_x$CuO$_4$ (LSCO) used in this Letter are within the VEPQL scheme  
the effective interaction and transfer integral ratio $U/4t\approx 1.525$ where $t\approx 295$ meV and
$T=0$ critical hole concentrations $x_c\approx 0.05$ and $x_*=0.27$ for both such systems,
lattice spacing $a\approx 3.9$ \AA, average separation between CuCO$_2$ planes $d_{\parallel}\approx 5.9$ \AA,
maximum $s1$ fermion pairing energy per spinon
$\Delta_0\approx 84$ meV, and coefficient $C_{s1}=1$ for YBCO 123 and $a\approx 3.8$ \AA, $d_{\parallel}\approx 6.6$ \AA,
$\Delta_0\approx 42$ meV, and $C_{s1}=2$ for LSCO \cite{cuprates0}. The VEPQL predictions achieve a good agreement with 
the cuprates universal properties \cite{cuprates0} and those of their parent compounds \cite{companion2}
and consistency with the coexisting two-gap scenario \cite{two-gaps}:
A pseudogap $2\vert\Delta\vert\approx (1-x/x_*)2\Delta_0$ and superconducting energy scale 
$2\vert\Omega\vert\approx 4k_B T_c/(1-[x_c/x_*](T_c/T_c^{max}])$ over the whole dome $x\in (x_c,x_*)$, 
where $T_c \approx \gamma_d\,[(x-x_c)/(x_*-x_c)](1-x/x_*)[\Delta_0/2k_B]$ and $(1-x_c/x_*)\geq\gamma_d\geq 1$. 
Those are the maximum magnitudes of the spinon pairing energy and superconducting virtual-electron pairing
energy, respectively. 

In the ground state there is no one-to-one correspondence between a $c$ fermion pair
and a two-spinon $s1$ fermion in that such objects may participate in several
virtual-electron pairs. Specifically, the strong effective coupling of $c$ fermion pairs whose hole
momenta $\vec{q}^{\,h}$ and $-\vec{q}^{\,h}$ belong to an approximately circular
$c-sc$ line centered at $-\vec{\pi}=-[\pi,\pi]$ results from interactions within virtual-electron pair configurations
with a well-defined set of $s1$ fermions whose two spinons momenta $\pm\vec{q}$ 
belong to a uniquely defined $s1-sc$ line arc centered at $\vec{0}=[0,0]$.
A $c-sc$ line has radius $q^h=\vert\vec{q}^{\,h}\vert\in (q^h_{Fc},q^h_{ec})$ and $c$ fermion 
energy $\vert\epsilon_c (q^h)\vert \in (0,W_{ec})$ such that $\vert\epsilon_c (q^h_{Fc})\vert=0$ and  
$\vert\epsilon_c (q^h_{ec})\vert=W_{ec}=4\Delta_0/(1-x_c/x_*)$. For Fermi angles $\phi\in (0,\pi/2)$
the corresponding $s1-sc$ line arc can be labelled either by its nodal momentum absolute
value $q^N_{arc} = q^N_{arc} (q^h)\in (q^N_{ec},q^N_{Bs1})$ or angular width
$2\phi_{arc} = \arcsin ([q_{arc}^N -q^N_{ec}]/[q^N_{Bs1}-q^N_{ec}])\in (0,\pi/2)$.
Here and above $q^N_{ec} \approx q^N_{Bs1} - [\Delta_0/t]\{C_{s1}/[(x_*-x_c)\sqrt{(q^{AN}_{Bs1})^2-(q^{N}_{Bs1})^2}]\}$,
$q^{h}_{ec} \approx  (\sqrt{1+[\Delta_0/ x (x_*-x_c)\pi^2 t}]) q_{Fc}^h$,
$q^N_{Bs1}$ and $q^{AN}_{Bs1}$
are nodal and anti-nodal momentum absolute values, respectively, belonging to the strongly anisotropic
$s1$ band boundary-line centered at $\vec{0}$ \cite{companion2}, and 
$q^h_{Fc}\approx \sqrt{x\,\pi}\,2$ refers to the isotropic $c$ Fermi line centered at $-\vec{\pi}$.
The energy needed for the $c$ fermion strong effective coupling is supplied by the short-range spin correlations through the $c$ - $s1$
fermion interactions within each virtual-electron pair configuration. 
{\it Strong} $c$ fermion effective coupling is that whose corresponding
virtual-electron pair breaking under one-electron removal excitations gives rise to sharp-feature-line arcs 
centered at momenta $\pm\vec{\pi}=\pm [\pi,\pi]$. Those are
in one-to-one correspondence to the $s1-sc$-line arcs of the virtual-electron pair $s1$ fermion. Such
sharp-feature-line arcs have angular range $\phi\in (\pi/4 -\phi_{arc},\pi/4 +\phi_{arc})$ and
energy $E \approx 2W_{ec} (1-\sin 2\phi_{arc})$. Hence they
exist only for $E < E_1 (\phi) = 2W_{ec}(1-\vert\cos 2\phi\vert)$. 
The macroscopic condensate refers to $c$ fermion pairs  
whose phases $\theta=\theta_0 +\theta_1$ line up. The fluctuations of $\theta_0$ and $\theta_1$ 
become large for $x\rightarrow x_c$ and $x\rightarrow x_*$, respectively. The dome $x$
dependence of the critical temperature $T_c$ is fully determined by the interplay of such fluctuations. 
A pseudogap state with short-range spin order and virtual-electron pair configurations without phase coherence occurs 
for temperatures $T\in (T_c,T^*)$ where $T^* \approx C_{s1}(1-x/x_*)[\Delta_0/2k_B]$ is the pseudgap temperature.
At $T=0$ a normal state emerges by application of a magnetic field aligned perpendicular to the planes of magnitude 
$H\in (H_0,H_{c2})$ for $x\in (x_0,x_{c2})$ and $H\in (H_0,H^*)$ for $x\in (x_{c2},x_*)$. 
The fields $H_0$, $H_{c2}$, and $H^*$ and the hole concentration $x_0<x_c$ are given in Ref. \cite{cuprates0}.
For $x\in (x_0,x_{c1})$ the upper magnetic field $H_{c2}(x)$ refers to the straight line plotted in Fig. 4 of that reference 
where $x_0\approx 0.013$ and $x_{c1}=1/8$. However, for $x\in (x_1,x_{c2})$ the actual 
$H_{c2}(x)$ line may (or may not) slightly deviate to below the straight line plotted in that figure. If so, the 
hole concentration $x_{c2}\approx 0.20$ may increase to $\approx 0.21-0.22$. Fortunately, 
such a possible deviation does not change the physics discussed here.

The main goals of this Letter are: i) The study of the one-electron scattering rate and normal-state $T$-dependent 
resistivity within the VEPQL; ii) Contributing to the further understanding of the role of scale-invariant physics in the unusual
scattering properties of the hole-doped cuprates. Our results refer to a range $x\in (x_A,x_{c2})$ 
for which $V_{Bs1}^{\Delta}/V_{Fc}\ll 1$. Here $x_A\approx x_*/2 =0.135$
and the $s1$ boundary line and $c$ Fermi velocities read $V_{Fc}\equiv V_c(\vec{q}^{\,h\,d}_{Fc}) 
\approx [\sqrt{x\pi}\,2/m_c^*]$ and $V^{\Delta}_{Bs1}\equiv V^{\Delta}_{s1} ({\vec{q}}^{\,d}_{Bs1}) 
\approx [\vert\Delta\vert/\sqrt{2}]\vert\sin 2\phi\vert$, respectively, where $m_c^*$ is the $c$ fermion mass.
For $x\in (x_{c2},x_*)$ that inequality is also fulfilled but there emerge competing scattering processes difficult to describe in 
terms of $c$ - $s1$ fermion interactions. Elsewhere it is shown that the VEPQL
predictions agree quantitatively with the distribution of the LSCO sharp photoemission 
spectral features of Figs. 3 and 4 of Ref. \cite{LSCO-ARPES-peaks}.
As predicted, they occur for energies $E (\phi)< E_1 (\phi)$ and the corresponding sharp-feature line arcs angular ranges
agree with the theoretical magnitudes.
This reveals experimental spectral signatures of the VEPQL virtual-electron pairing mechanism.

Here we start by using a Fermi's golden rule in terms of the $c$ - $s1$ fermion interactions to 
calculate for small $\hbar\omega$ the one-electron inverse lifetime. Upon removal of one electron, 
two holes emerge in the $s1$ and $c$ bands, respectively. For low transfer energy $\hbar\omega$ 
and small transfer momentum $\vec{p}$ the $c$ - $s1$ fermion inelastic collisions conserve  
the doublicity $d=\pm 1$, which refers to one-electron excited states with the same energy
and momentum but different electron velocity \cite{companion2,cuprates0}. 
Within such processes one $s1$ fermion moves to the single hole in the $s1$ band. 
One must then integrate over all particle-hole or hole-particle processes 
in the $c$ fermion band that conserve energy and momentum. 
For low $\hbar\omega$ and small $\vec{p}$ the one-electron inverse lifetime
can then be written as,
\begin{widetext}
\begin{equation}
{\hbar\over\tau_{el,d}} = 2\pi\int{d{q^h}^2\over [2\pi]^2}
\vert W_{c,s1} ({\vec{q}}^{\,h},{\vec{q}}^{\,d}_{Bs1};{\vec{p}})\vert ^2\,
N_c({\vec{q}}^{\,h})N^h_c({\vec{q}}^{\,h}+\vec{p})
N^h_{s1}({\vec{q}}^{\,d}_{Bs1}-\vec{p})\,\delta (\epsilon_c ({\vec{q}}^{\,h}+\vec{p})-\epsilon_c ({\vec{q}}^{\,h})
+\epsilon_{s1} ({\vec{q}}^{\,d}_{Bs1}-\vec{p})-\epsilon_{s1} ({\vec{q}}^{\,d}_{Bs1})) \, ,
\label{tau}
\end{equation}
\end{widetext}
where $d=\pm 1$. The number of $c$ fermions equals that of spin-up plus spin-down electrons, so that there 
is no additional factor $2$ in this expression.
The $c$ and $s1$ fermion energy dispersions and momentum distribution functions appearing here are introduced in Refs. \cite{companion2,cuprates0}
and $W_{c,s1} ({\vec{q}}^{\,h},{\vec{q}};{\vec{p}})$ is the matrix element of 
the $c$ - $s1$ fermion effective interaction between the initial and final states. 
It can be estimated for the hole concentration range $x\in (x_A,x_{c2})$ for which
$r_{\Delta} =V^{\Delta}_{Bs1}/V_{Fc}\ll 1$. Indeed, then the single heavy $s1$ fermion 
hole plays mainly the role of a scattering center and the $c$ fermion holes that of scatterers 
and one can evaluate the matrix element absolute value 
$\vert W_{c,s1} ({\vec{q}}^{\,h},{\vec{q}};{\vec{p}})\vert$ to zeroth order in $V^{\Delta}_{Bs1}/V_{Fc}\ll 1$. 
Provided that the small velocity $V^{\Delta}_{Bs1}$ is accounted for in the physical quantities 
of expression (\ref{tau}) other than that matrix element, such a procedure leads to a good approximation for 
the one-electron inverse lifetime $\hbar\omega$ dependence. For that $x$ range we then find
$\lim_{{\vec{p}}\rightarrow 0}\vert W_{c,s1} ({\vec{q}}^{\,h},{\vec{q}};{\vec{p}})\vert \approx 
[\pi/4\rho_c (\vec{q}^{\,h})] \vert\sin (\delta_1 -\delta_0)\vert$ where $\vec{q}^{\,h}\approx\vec{q}^{\,h\,d}_{Fc}$, 
$\vec{q}\approx{\vec{q}}_{Bs1}^{\,d}$, the angular-momentum
$l=0,1$ phase shifts read $\delta_{0} = \pi/2$ and $\delta_{1} = 2\phi$, respectively, and
the density of states $\rho_c (\vec{q}^{\,h}) = m_c^*/2\pi\hbar^2$ of the present effective
two-dimensional $c$ fermion scattering problem is independent of $x$. Hence we arrive to
$\lim_{{\vec{p}}\rightarrow 0}\vert W_{c,s1} ({\vec{q}}^{\,h},{\vec{q}};{\vec{p}})\vert\approx 
[\pi^2\hbar^2/2m_c^*]\vert\cos 2\phi\vert\approx \pi^3 x_* t \vert\cos 2\phi\vert$. Fortunately,
for $x\in (x_A,x_{c2})$ the quantities contributing to (\ref{tau}) are independent
of the doublicity $d=\pm 1$, so that after some algebra we arrive to an inverse lifetime
$\hbar/\tau_{el}\approx \hbar\omega\,\pi\alpha_{\tau_{el}}$ and
scattering rate $\Gamma (\phi,\omega) =1/[\tau_{el} \,V_F]\approx \hbar\omega\,\pi\alpha$. 
Here $V_F\approx V_{Fc}\approx [\sqrt{x\pi}\,2/m_c^*]$, $\alpha_{\tau_{el}} = [\pi/16 \sqrt{x\,x_{op}}](\cos 2\phi)^2$, and 
$\alpha = (\cos 2\phi)^2/(x\,64x_*\sqrt{\pi x_{op}}\,t)$ where $x_{op}=(x_*+x_c)/2=0.16$. (We use units
of lattice constant $a=1$.) Such small-$\hbar\omega$ expressions 
are expected to remain valid for approximately $\hbar\omega <E_1 (\phi)= 2W_{ec}(1-\vert\cos 2\phi\vert)$.
The factor $(\cos 2\phi)^2$ also appears in the anisotropic component
of the scattering rate studied in Ref. \cite{scattering-rate} for hole concentrations $x>x_{c2}$.
For $x=0.145$ the use of the LSCO parameters leads to the theoretical 
coefficient $\alpha (\phi)$ plotted in Fig. \ref{fig1} (solid line) 
together with the experimental points
of Fig. 4 (c) of Ref. \cite{PSI-ANI-07} for $[\alpha_I (\phi)-\alpha_I (\pi/4)]$.
(The very small $\alpha_{I} (\pi/4)$ magnitude is related 
to processes that are not contained in the VEPQL.)
An excellent quantitative agreement is obtained  
between $\alpha (\phi)$ and the experimental points of  
$[\alpha_I (\phi)-\alpha_I (\pi/4)]$. 
\begin{figure}
\includegraphics[width=3.75cm,height=3.75cm]{resist-L-new-sirius.eps}
\includegraphics[width=3.75cm,height=3.75cm]{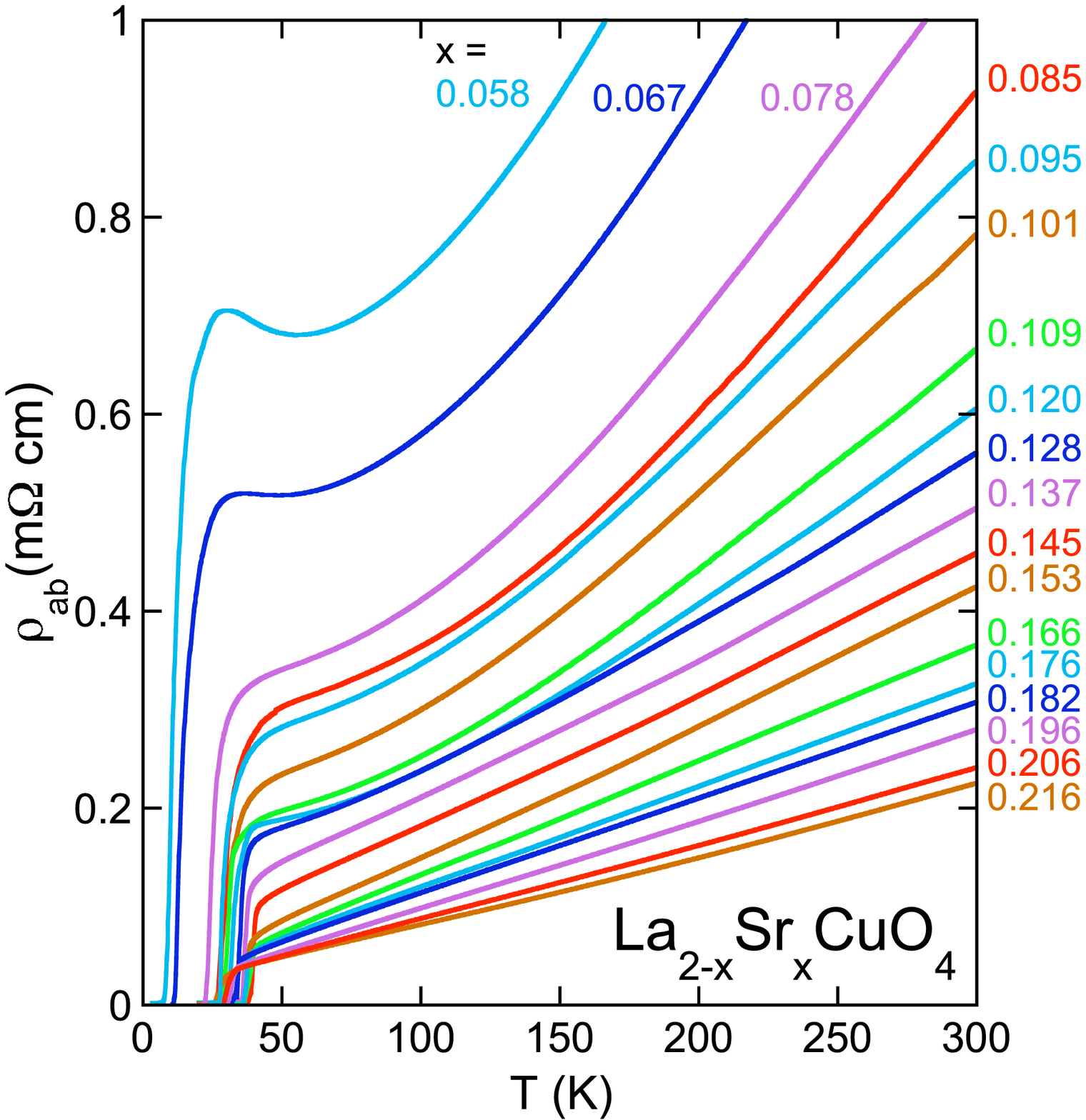}
\caption{(a) The $T$ dependence of the resistivity
$\rho (T,0)=\theta (T-T_c)\,\rho (T)$ with
$\rho (T)$ given in Eq. (\ref{rho-F}) for $x\in (x_A,x_{c2})$
where $x_A\approx 0.135$ and $x_{c2}\approx 0.20-0.22$ and the parameter 
magnitudes for LSCO. (b) Corresponding experimental curves.
Experimental curves figure from Ref. \cite{L-resistivity}.}
\label{fig2} 
\end{figure}

In the following we provide strong evidence from agreement between theory
and experiments that the linear-$T$ resistivity is indeed 
a manifestation of normal-state scale-invariant physics. 
This requires that the $T$-dependence of the inverse relaxation lifetime 
derived for finite magnetic field, $x\in (x_A,x_{c2})$, and
$\hbar\omega\ll \pi k_B\,T$ by replacing $\hbar\omega$ by $\pi k_B\,T$ in the  
one-electron inverse lifetime $1/\tau_{el}$ and averaging over the Fermi line
leads to the observed low-$T$ resistivity. To access the low-$T$ resistivity for the normal state 
a magnetic field perpendicular to the planes is applied, which
remains unaltered down to $T=0$, as in the cuprates \cite{critical-2}.
The field serves merely to remove superconductivity and achieve the $H$-independent 
term $\rho (T)$ of $\rho (T,H) = \rho (T)+\delta\rho (T,H)$
where $\delta\rho (T,H)$ is the magnetoresistance contribution.
The $T$-dependent inverse relaxation lifetime 
derived by replacing $\hbar\omega$ by $\pi k_B\,T$ in the 
above one-electron inverse lifetime $\hbar/\tau_{el}\approx\hbar\omega\,\pi\alpha_{\tau_{el}}$ 
and averaging over the Fermi line is given by,
\begin{eqnarray}
{1\over\tau (T)} & = & {2\over\pi}\left(\int_{0}^{\pi/2}d\phi{1\over\tau_{el}}
\right)\Big\vert_{\hbar\omega=\pi k_B T} = 
{1\over\hbar A}\,\pi k_B T  \, ,
\nonumber \\
A & = & {32\over \pi^2}\sqrt{x\,x_{op}} \, , \hspace{0.25cm} x \in (x_A,x_{c2}) \, .
\label{1-over-tau}
\end{eqnarray}
The hole concentration $x_A\approx x_*/2$ is that at
which $A\approx 0.5$ becomes of order one. The normal-state
resistivity $H$-independent term $\rho (T)$ of $\rho (T,H)$ then reads,
\begin{equation}
\rho (T) \approx \left({m_c^{\rho}\,d_{\parallel}\over x e^2}\right){1\over\tau (T)} 
\, ; \hspace{0.25cm} m_c^{\rho} = {\hbar^2 \pi x_*\over 2t} \, ,
\label{rho-F-tau}
\end{equation}
where $m_c^{\rho}$ is the $c$ fermion transport mass \cite{companion2}. Combination of Eqs. (\ref{1-over-tau}) 
and (\ref{rho-F-tau}) leads to the following resistivity expression,
\begin{equation}
\rho (T) \approx \left({\hbar \,d_{\parallel}\over t e^2}\right)\left({\pi\over 4}\right)^3{x_*\over x^{3/2}\sqrt{x_{op}}}\,\pi k_B T  \, .
\label{rho-F}
\end{equation}
Consistency with the above $\hbar/\tau_{el}$ expression validity range
$\hbar\omega <E_1 (\phi)$ implies that the behavior (\ref{rho-F}) remains dominant 
in the normal-state range $T\in (0,T_1)$. Here, 
\begin{equation}
T_1\approx {2\over\pi}\int_{0}^{\pi/2}d\phi {E_1(\phi)\over k_B}= {(2\pi-4)W_{ec}\over \pi^2 k_B} \, .
\label{T-1}
\end{equation}
At $x=0.16$ this gives $T_1\approx 554$\,K for LSCO and $T_1\approx 1107$\,K for YBCO 123.
Extrapolation of expression (\ref{rho-F}) to $H=0$ 
leads to $\rho (T,0)\approx \theta (T-T_c) \rho (T)$ for $T<T_1$. Here
the critical low-$T$ resistivity behavior (\ref{rho-F}) is masked by the 
onset of superconductivity at $T=T_c$. 

\begin{figure}
\includegraphics[width=3.75cm,height=3.75cm]{resist-new-new.eps}
\includegraphics[width=3.75cm,height=3.75cm]{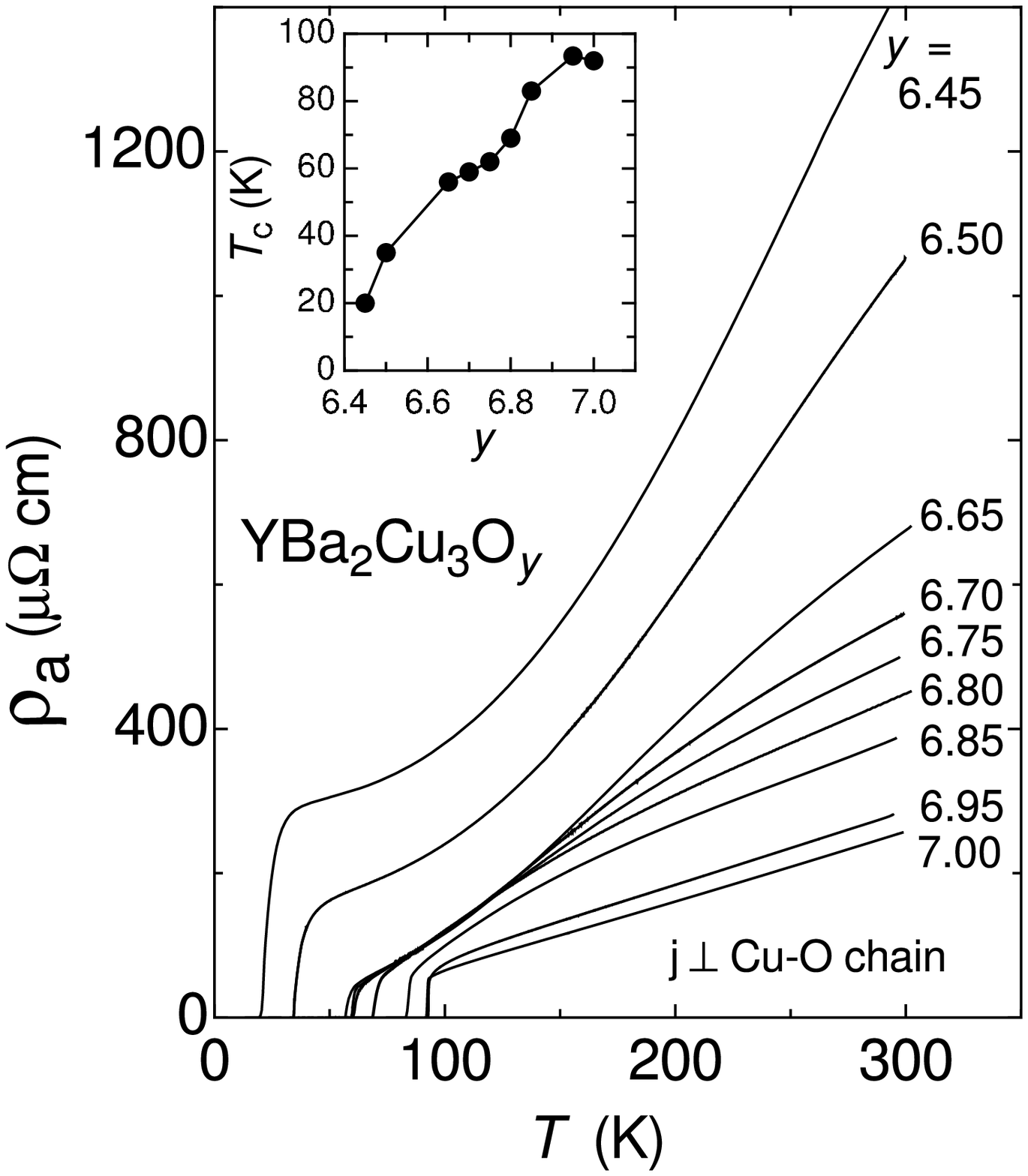}
\caption{(a) The $T$ dependence of the resistivity
$\rho_a (T,0)=\theta (T-T_c)\,\rho (T)$ where $\rho (T)$
is given in Eq. (\ref{rho-F}) for the parameter 
magnitudes for YBCO 123 for a set of $y$ values. 
(b) Corresponding experimental curves of Ref. \cite{Y-resistivity}.
The oxygen content $y-6$ is obtained from
$x$ by use of Fig. 4 (a) of Ref. \cite{Tc-1/8}. The theoretical $T_c$ is 
larger at $y=6.95$ than for $y=6.85,7.00$. This is alike in the inset of the 
second figure. Experimental curves figure from Ref. \cite{Y-resistivity}.}
\label{fig3} 
\end{figure}

We now compare our theoretical linear-$T$ resistivity with that of LSCO \cite{L-resistivity} and
YBCO 123 \cite{Y-resistivity} for $H=0$ and $T$ up to $300$\,K.
Transport in the $b$ direction has for YBCO 123
contributions from the CuO chains, which render our results unsuitable. In turn,
$\rho_a (T,0)\approx \theta (T-T_c) \rho (T)$ at $H=0$ for the $a$ direction. 
$\rho (T,0)$ and $\rho_a (T,0)$ are plotted in Figs. \ref{fig2} and
\ref{fig3} for the parameters for LSCO and YBCO 123, respectively. $x$  
is between $x\approx x_A\approx 0.135$ and $x_{c2}\approx 0.20-22$ for the 
LSCO theoretical lines of Fig. \ref{fig2}. Fig. \ref{fig3} for YBCO 123 refers to three $x$ values near 
$x_{op}$, expressed in terms of the oxygen content.
Comparison of the theoretical curves of Fig. \ref{fig2}
with the LSCO resistivity curves of Ref. \cite{L-resistivity} also shown
in the figure confirms an excellent quantitative agreement
between theory and experiments for the present range $x\in (x_A,x_{c2})$. 
In turn, for YBCO 123 our scheme provides a good quantitative description of the experimental curves
near $x_{op}$, for $y=6.95,7.00$. The $y=6.85$ experimental curve of Ref. \cite{Y-resistivity} 
already deviates from the linear-$T$ behavior. (The hole concentration that
marks the onset of such a behavior reads for that material $x_A\approx 0.15$.)

For the present range $x\in (x_A,x_{c2})$, the interplay of the $c$ Fermi line isotropy 
with the $s1$ boundary line strong anisotropy \cite{companion2,cuprates0} 
plays an important role. It is behind the $c$ - $s1$ fermion inelastic collisions 
leading to anisotropic one-electron scattering properties associated with
the factor $(\cos 2\phi)^2$ in the one-electron scattering rate expression. 
In turn, consistently with the experimental resistivity curves of Figs. \ref{fig2} and
\ref{fig3}, the non-linear $T$ 
dependence of the resistivity developing for approximately $x<x_A$ for a range of
low temperatures that increases upon decreasing $x$ is in part due to 
the matrix element $W_{c,s1} ({\vec{q}}^{\,h},{\vec{q}};{\vec{p}})$ acquiring 
a different form due to the increase of the ratio $r_{\Delta} =V^{\Delta}_{Bs1}/V_{Fc}$ magnitude. 
Our method does not apply to that regime.
On the other hand, for the range $x>x_{c2}$ also not considered here a competing scattering channel 
emerges, leading to an additional $T^2$-quadratic resistivity contribution \cite{critical-2,Resistivity}. 

That the dependence on the Fermi angle  $\phi\in (0,\pi/2)$ of the scattering-rate
coefficient $\alpha=\alpha_{\tau_{el}}/\hbar V_F=(\cos 2\phi)^2/(x\,64x_*\sqrt{\pi x_{op}}\,t)$
associated with that of the inverse lifetime $\hbar/\tau_{el}\approx \hbar\omega\,\pi\alpha_{\tau_{el}}$ 
agrees with the experimental 
points of Fig. \ref{fig1} seems to confirm the evidence provided in
Refs. \cite{companion2,cuprates0} that the VEPQL may contain some of
the main mechanisms behind the unusual properties of the hole-doped cuprates and
their parent compounds.
That in addition the inverse relaxation lifetime $1/\tau$ $T$-dependence 
obtained for $\hbar\omega\ll \pi k_B\,T$ and $x\in (x_A,x_{c2})$
by merely replacing $\hbar\omega$ by $\pi k_B\,T$ in $1/\tau_{el}$ 
and averaging over the Fermi line leads to astonishing quantitative agreement with 
the resistivity experimental lines is a stronger surprising result. 
Consistently, for $\hbar\omega\ll \pi k_B\,T$ and $x\in (x_A,x_{c2})$ the system exhibits dynamics 
characterized by the relaxation time $\tau = \hbar A/\pi k_B\,T$ of Eq. (\ref{1-over-tau}), 
where $A = [32\sqrt{x\,x_{op}}/\pi^2]\approx 1$ for $x> x_A$. 
This second stronger result provides clear evidence
of normal-state scale-invariant physics. It may follow from beyond mean-field theory the $T=0$ 
line $H_{c2}(x)$, plotted in Fig. 4 of Ref. \cite{cuprates0} for $x\in (x_0,x_{c2})$, referring to a true quantum phase transition.
Such a transition could occur between a state with long-range spin order 
regulated by monopoles and antimonopoles for $H>H_{c2}$ and a vortex liquid  
by vortices and antivortices for $H<H_{c2}$. 
That would be a generalization for $H_{c2}>0$ of the quantum phase transition speculated to occur 
at $x_{c2}\approx x_0$ and $H_{c2}\approx 0$ in Ref. \cite{duality}. The field $H^{*}$ 
marks a change from a short-range spin order to a disordered state and thus refers to a crossover.
Hence the normal-state scale invariance occurring for $x\in (x_A,x_{c2})$ could result from 
the hole concentration $x_{c2}\approx 0.2$, where the lines of Fig. 4 of Ref. \cite{cuprates0}
associated with the fields $H_{c2} (x)$ and $H^* (x)$ meet, referring to a quantum critical point
\cite{LSCO-1/lam-x,critical}. Such a quantum critical point may prevent the competing 
$T^2$-quadratic resistivity contribution to strengthen below $x=x_{c2}$.

\begin{acknowledgments}
I thank M. A. N. Ara\'ujo, J. Chang, A. Damascelli, R. Dias, P. Horsch, S. Komiya, 
J. Mesot, C. Panagopoulos, T. C. Ribeiro, P. D. Sacramento, M. J. Sampaio, M. Shi, and 
Z. Te${\rm\check{s}}$anovi\'c for discussions, the authors of Refs. \cite{PSI-ANI-07,L-resistivity,Y-resistivity}
for providing the experimental data of Figs. 1-3,
and the support of the ESF Science Program INSTANS and grant PTDC/FIS/64926/2006.
\end{acknowledgments}


\begin{references}
\bibitem{LSCO-1/lam-x} 
	C. Panagopoulos {\it et al.}, Phys. Rev. B {\bf 66}, 064501 (2002).
\bibitem{critical} 
	D. van der Marel {\it et al.}, Nature {\bf 425}, 271 (2003).
\bibitem{critical-2} 
	R. Daou {\it et al.} Nature Phys. {\bf 5}, 31 (2009).
\bibitem{Resistivity}           
     	R. A. Cooper {\it et al.}, Science {\bf 323}, 603 (2009).
\bibitem{QPT} 
	R. Jaramillo {\it et al.}, Nature {\bf 459}, 405 (2009).
\bibitem{ARPES-review} 
	A. Damascelli, Z. Hussain, and X. Z. Shen, Rev. Mod. Phys. {\bf 75}, 473 (2003).
\bibitem{2D-MIT} 
	P. A. Lee, N. Nagaosa, and X. G. Wen,  Rev. Mod. Phys. {\bf 78}, 17 (2006).
\bibitem{duality}
	Z. Te${\rm\check{s}}$anovi\'c, Nature Phys. {\bf 4}, 408 (2008).
\bibitem{symmetry}
	J. M. P. Carmelo, Stellan \"Ostlund, and M. J. Sampaio, Annals Phys. (2010), doi: 10.1016/j.aop.2010.03.002.
\bibitem{companion2}
	J. M. P. Carmelo, Nucl. Phys. B {\bf 824}, 452 (2010).
\bibitem{cuprates0}
	J. M. P. Carmelo, arXiv:1004.0923 (2010).
\bibitem{two-gaps}
	S. H\"ufner, M. A. Hussain, A. Damascelli, and G. A. Sawatzky, 
	Rep. Prog. Phys. {\bf 71}, 062501 (2008).
\bibitem{Yu-09} 	
	G. Yu, Y. Li, E. M. Motoyama, and M. Greven, Nature Phys. {\bf 5}, 873 (2009).
\bibitem{k-r-spaces}
	Y. Kohsaka {\it et al.}, Nature Phys. {\bf 5}, 642 (2009).
\bibitem{PSI-ANI-07}
	J. Chang {\it et al.}, Phys. Rev. B {\bf 78}, 205103 (2008).
\bibitem{two-gap-Bi}           
     	M. C. Boyer {\it et al.}, Nature Phys. {\bf 3}, 802 (2007).
\bibitem{Hm}           
     	L. Li, J. G. Checkelsky, S. Komiya, Y. Ando, and N. P. Ong, 
     	Nature Phys. {\bf 3}, 311 (2007).
\bibitem{LSCO-ARPES-peaks} 
	J. Chang {\it et al.}, Phys. Rev. B {\bf 75}, 224508 (2007).
\bibitem{scattering-rate}
	M. Abdel-Jawad {\it et al.}, Nature Phys. {\bf 2}, 821 (2006).
\bibitem{Pines} 
	D. Pines and P. Nozi\`eres, in {\it The theory of quantum liquids}
	(Benjamin, New York, 1996) Volume 1.
\bibitem{L-resistivity}
	S. Komiya, H.-D. Chen, S. C. Zhang, and Y. Ando, Phys. Rev. Lett. {\bf 94}, 207004 (2005).
\bibitem{Y-resistivity} 
	K. Segawa and Y. Ando, Phys. Rev. Lett. {\bf 86}, 4907 (2001).
\bibitem{Tc-1/8}
	R. Liang, D. A. Bonn, and W. N. Hardy, Phys. Rev. B {\bf 73}, 180505 (2006).
\end{references}
\end{document}